\documentclass[12pt]{article}
\usepackage{epsf}
\def\beq{\begin{equation}}
\def\eeq{\end{equation}}
\def\bea{\begin{eqnarray}}
\def\beaa{\begin{eqnarray*}}
\def\eea{\end{eqnarray}}
\def\eeaa{\end{eqnarray*}}
\def\bq{\begin{quote}}
\def\eq{\end{quote}}
\def\gappeq{\mathrel{\rlap {\raise.5ex\hbox{$>$}} {\lower.5ex\hbox{$\sim$}}}}

\def\lappeq{\mathrel{\rlap{\raise.5ex\hbox{$<$}} {\lower.5ex\hbox{$\sim$}}}}

\begin{document} 
\begin{titlepage}
\pagestyle{empty}
\baselineskip=21pt
\rightline{hep-ph/0012067}
\rightline{CERN--TH/2000-358}
\vskip 1in
\begin{center}
{\large \bf A light Higgs Boson would invite Supersymmmetry}\\
\vspace*{1cm}
{\bf John Ellis} and {\bf Douglas Ross}
\footnote{On leave of absence from: {\it
 Department of Physics and Astronomy, University of Southampton,
 Southampton SO17 1BJ, UK}}
  \medskip

 {\it TH Division, CERN, Geneva, Switzerland}

\vskip 1in
{\bf Abstract}
\end{center}
\baselineskip=18pt \noindent

If the Higgs boson weighs about 115 GeV, the effective potential of the
Standard Model becomes unstable above a scale of about $10^6$~GeV. This
instability may be rectified only by new {\it bosonic} particles
such as stop squarks. 
However, avoiding the instability requires fine-tuning of the model
couplings, in particular if the theory is not to become non-perturbative
before the Planck scale. Such fine-tuning is automatic in a supersymmetric
model, but is lost if there are no Higgsinos. A light Higgs boson would be
{\it prima facie} evidence for
supersymmetry in the top-quark and Higgs sectors. 

\vfill
\vskip 0.15in
\leftline{CERN--TH/2000-358}
\leftline{December 2000}
\end{titlepage}
\baselineskip=18pt

\section{Introduction}

The LEP collaborations~\cite{heroes} and the LEP Working Group for Higgs
boson searches~\cite{HWG} have recently reported excesses of events that
might be due to the production of a Higgs boson weighing about 115~GeV.
Confirmation of this exciting interpretation of their events may have to
wait for several years, but we already find the hint sufficiently
encouraging to explore the possible implications of such a discovery. As
has already been pointed out~\cite{EGNO}, the existence of such a light
Higgs boson would imply that the Standard Model could remain valid only up
to
scales limited by about $10^6$~GeV~\cite{AI}. This is because the
effective Higgs potential would be destabilised by the radiative
corrections due to the relatively heavy top quark, that could not be
counterbalanced by those due to a relatively light Higgs boson alone,
necessitating the appearance of new physics. Extant non-perturbative
models of new physics, such as technicolour, cannot accommodate a
relatively light Higgs boson~\cite{TC}. On the other hand, 
a relatively light Higgs boson favours {\it prima facie} generically
a perturbative scenario for electroweak symmetry breaking.
Among these, it is well
known that a Higgs boson weighing less than about 130~GeV is expected in
the minimal supersymmetric extension of the Standard Model
(MSSM)~\cite{EFZ}. 

The question we address in this paper is the extent to which a Higgs boson
weighing about 115~GeV would actually {\it require} significant features
of supersymmetry. Salient features of the MSSM that one might look for
include (a) bosonic partners for known fermions, and vice versa, (c) the
absence of certain renormalization effects, entailed in the MSSM by
relations between bosonic and fermionic couplings, and (c) soft
supersymmetry breaking at a scale of about $10^3$~GeV. 

Remarkably, we find that a 115~GeV Higgs boson would provide non-trivial
hints for all these characteristics of the MSSM. Specifically, analyzing 
the renormalization of the effective potential, we find
that (a) the new physics that must appear at some energy scale below
$10^6$~GeV has to contain a dominant bosonic component, just as provided
by the stop squarks in the MSSM, (b) that this new component must be
finely tuned, just as occurs when supersymmetry relates the couplings of
bosons and fermions in the top and Higgs supermultiplets,  (c) that the
splitting of
bosonic and fermionic partners should be in the range $10^2$ to
$10^5$~GeV. Further, analyzing precision electroweak observables, we
show (d) that the effective mass splitting between members of a scalar
isomultiplet must be small, as happens in the MSSM.
The new physics may not be supersymmetric, but it must  share many
features with supersymmetry.

\section{Renormalization of the Effective Potential}

We first consider in more detail the renormalization of the effective
Higgs potential. In the minimal Standard Model, there are important
renormalization effects of the quartic Higgs self-coupling $\lambda_H$ due
to the top-quark Yukawa coupling $g_t$, as well as those due to
$\lambda_H$ itself \cite{cjse}:
\beq  \beta_{\lambda_H} \ \equiv
 \ \frac{\partial}{\partial \mu} \lambda_H
\ = \ \frac{1}{16 \pi^2} \left( 4 \lambda_H^2 \, - \, 36 \, g_t^4
 \, +\frac{27}{4} g^4 \, + \, \frac{9}{4} g^{\prime \, 4}
+ \, \frac{9}{2} g^2 g^{\prime \, 2} \right) + 2 \lambda_H \gamma_H,
 \label{bethsm}
 \eeq
where $\gamma_H$ is the one-loop wave-function renormalization
constant of the Higgs  field:
\beq
\gamma_H \ = \ \frac{1}{16\pi^2} \left( 6 g_t^2-\frac{9}{2} g^2 \ - 
\frac{3}{2} g^{\prime \, 2} \right),
\label{wfr}
\eeq
where $g$ and $g^\prime$ are the  
$SU(2)$ and $U(1)$ gauge couplings respectively. We neglect all the other
Yukawa couplings.  

The effective potential of the Standard Model becomes unstable at a value
of the Higgs field $H$ very close to that where $\lambda_H$ turns
negative. With $g_t$ fixed so that $m_t = 175$~GeV and $\lambda_H$
fixed so that $m_H = 115$~GeV, this instability scale
is no more than $10^6$~GeV, because the relatively large value of
$g_t$, which tries to decrease $\lambda_H$, overwhelms the
relatively small value of $\lambda_H$ itself. Thus, as already mentioned, 
new physics is needed to stabilise our electroweak vacuum. 

What form might this new physics take? Looking at the signs in
(\ref{bethsm}), it is clear that adding new fermions, e.g., in a fourth
generation, would only exacerbate the situation.  Only new {\it bosonic}
physics can fit the bill. The symmetry properties of this new physics are
not highly constrained, {\it a priori}, and one could imagine introducing
combinations of $N_{I,Y}$ new bosonic multiplets of isospin $I$ and weak
hypercharge $Y$. We recall that the MSSM contains an $N_{1/2, - 1/6} = N_C
= 3$ colour-triplet isospin doublets of left-handed stop and sbottom
squarks $({\tilde t}_L, {\tilde b}_L)$ and $N_{0, 2/3} = N_{0, -1/3} = N_C
= 3$ isospin singlets ${\tilde t}_R$ and ${\tilde b}_R$, but we shall not
postulate such a combination at the outset. Moreover, in the MSSM all the
quartic couplings are strictly related to $g_t$ and the 
gauge couplings, but we shall also not postulate such relations {\it a
priori}. 

To begin with, we consider the most general addition of
$n$ (complex) multiplets of 
scalar particles which each carry the same weak isospin $I$ and weak
hypercharge $Y$. 
These scalar particles are assumed {\it not} to partake in the
symmetry-breaking mechanism and so do not possess v.e.v.'s.
Denoting these scalars by $\phi_i^\alpha$,
where $i$ is the third component of isospin and $\alpha$ runs over the
number of copies, this gives rise to extra terms in the Lagrangian
\bea {\mathcal L}^\phi & = &  | D_\mu \phi^\alpha |^2 \, - \, 
 \left(M^2-M_0^2 \right)  | \phi^\alpha |^2 \,
 - \frac{2M_0^2}{v^2} |H|^2  | \phi^\alpha |^2 \, 
 - \frac{4(\Delta M^2)}{v^2} 
 \left( H^\dag \tau^a H  \right)  
 \left( \phi^{\dag \,\alpha} T^a \phi^\alpha \right)
\nonumber \\  & - & \frac{\lambda_\phi}{6} \left(| \phi^\alpha |^2 \right)^2
, \label{lagphi}
\eea
where the $\tau^a$ are the defining representation of the $SU(2)$
generators, the $T^a$ are the isospin-$I$ representation and $v$ is the
v.e.v. of the Higgs field.
We see in (\ref{lagphi}) that there are three contributions to the
masses of the added scalar particles - a conventional quadratic term and
two
contributions from couplings to the Higgs field. The quadratic term, $M$,
 sets
the overall scale of the new bosonic physics, and generalizes the soft
supersymmetry-breaking terms postulated in the MSSM. 
One of the Higgs quartic contributions in (\ref{lagphi}) conserves
custodial $SU(2)$ and contributes an average mass $M_0$ to the
multiplet, and the other breaks $SU(2)$, splitting the squared
masses of members of the multiplet with adjacent values of $I_3$ by
$(\Delta M^2)$. This term may have either sign.

\section{Stability of the Effective Potential}

The added scalar particles make a contribution to the one-loop
$\beta$ function for $\lambda_H$ (above the threshold $\mu=M$) of the form
\beq \Delta \beta_{\lambda_H} 
\ = \ \frac{6n}{4 \pi^2 v^4} \left((2I+1) M_0^4+(\Delta M^2)^2
 \frac{I(I+1)(2I+1)}{3} \right)  \label{dbetlam} \eeq 
Before one can insert this into the R.G.E. equation for $\lambda_H$, one
needs the $\beta$ functions for $M_0^2$ and $(\Delta M^2)$, which, at 
the one-loop level, are given by:
\bea \beta_{M_0^2} &= & \frac{M_0^2}{16\pi^2}
\left\{ 2 \lambda_H \, + \, \frac{4n}{3}(I+1) \lambda_{\phi}
\, + \, 8 \, \frac{M_0^2}{v^2} 
\, + \, \frac{3}{2} \left(g^4 I(I+1) + g^{\prime \, 4} Y^2 \right)
 \right\}
\nonumber \\ &+& M_0^2 \left(
 \, \gamma_H \, + \, \gamma_\phi \right), \label{dbm0} \eea
\bea \beta_{(\Delta M_0^2)} &= & \frac{(\Delta M^2)}{16\pi^2}
\left\{ 2 \lambda_H \, + \, \frac{4n}{3}(I+1) \lambda_{\phi}
\, + \, 8 \, \frac{M_0^2}{v^2}
\, - \, 3 \, Y  g^2  g^{\prime \, 2} 
  \right\}
\nonumber \\ &+& (\Delta M^2) \left(
 \, \gamma_H \, + \, \gamma_\phi \right), \label{dbdm} \eea
where $\gamma_\phi$ is the one-loop wave-function renormalization
constant of the scalar field,
\beq
\gamma_\phi \ = \ \frac{1}{16\pi^2} \left(-6 \, I(I+1) g^2 \ - 
6 \, Y^2 g^{\prime \, 2} \right).
\label{gammaphi}
\eeq
Finally, the one-loop $\beta$ function for the quartic self-coupling
$\lambda_\phi$ is
\bea \beta_{\lambda_\phi} & = & 
\frac{1}{16\pi^2}\left\{
48 \frac{M_0^4}{v^4} \, + \, \frac{2}{3} \lambda_\phi^2 
\left( (2I+1) n + 4 \right) + 3 g^4 I(I+1)(4I^2+6I-1) + 
36 \, g^{\prime \, 4} Y^4 \right. \nonumber \\ &+& \left.
  9 \,  g^2 g^{\prime \, 2} (2I+1) Y^2
   \right\}
\, + \,
  2 \, \lambda_\phi \gamma_\phi .
\label{betphi} \eea
The resulting set of coupled differential equations can readily be
solved by numerical methods.

We note that, as required, the extra contributions to the renormalization
of the Higgs quartic coupling are positive, thereby counteracting the
large contribution in (\ref{bethsm}) from the large top-quark Yukawa
coupling $g_t$. However, the effect of these extra terms is sensitive to
the
quartic self-coupling, $\lambda_\phi$, since this drives the running
of $M_0^2$ and $(\Delta M^2)$, which in turn drive the running of $\lambda_H$.
It would be unnatural for such a coupling to be too small.

\begin{figure}
\begin{center}
\leavevmode
\hbox{\epsfxsize=12 cm
\epsfbox{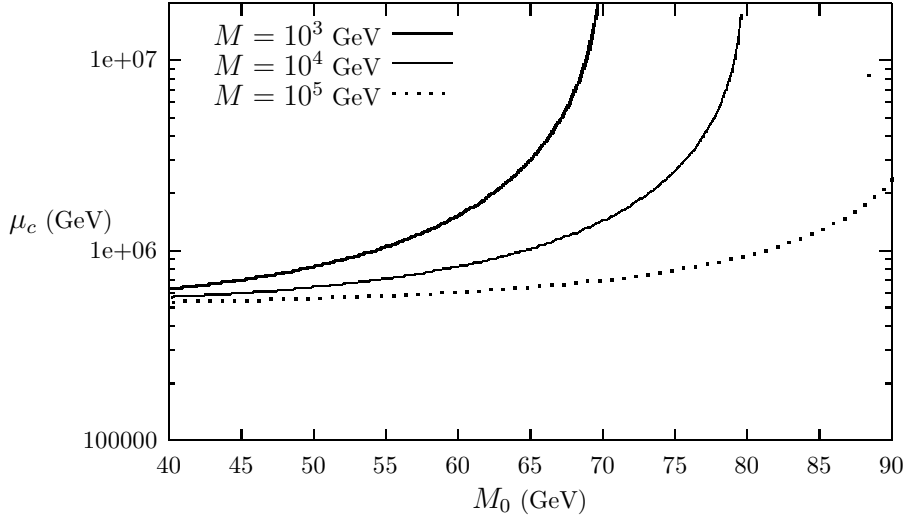}}
\vspace{1cm}
\caption{\it Examples of the increase in the critical scale $\mu_c$
at which the effective potential becomes unstable, as the 
coupling $M_0$ between the added scalar field $\phi$ and the Higgs field
$H$ is
increased, for three different choices of the scale $M$ of new physics.}
\label{fig1}
\end{center}
\end{figure}

\begin{figure}
\begin{center}
\leavevmode
\hbox{\epsfxsize=12 cm
\epsfbox{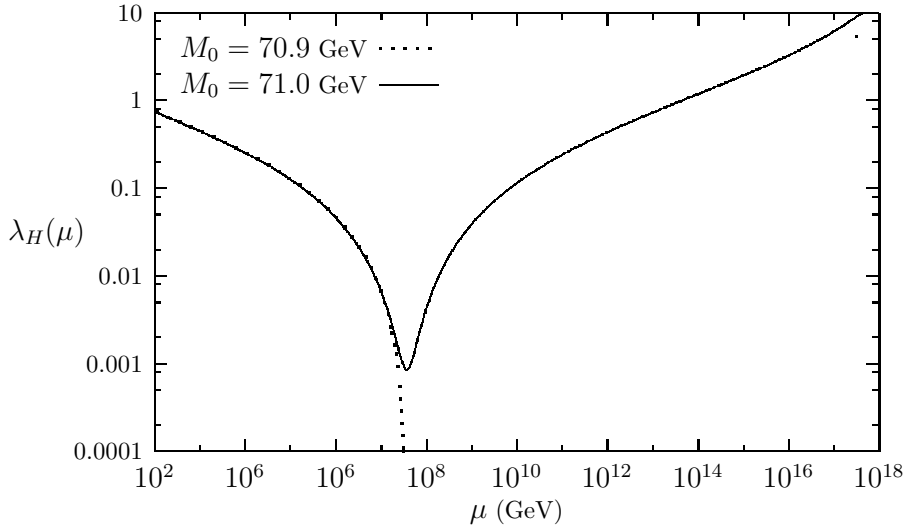}}
\vspace{1cm}
\caption{\it An example of the acute fine-tuning of $M_0$ that arises
when one
attempts to couple the added scalars to the Higgs fields in such a way
that the effective potential remains stable, whilst at the same
time remaining within the perturbative regime. Here the new physics
scale $M$ is taken to be 1 TeV.}
\label{fig2}
\end{center}
\end{figure}

As an example, we have taken the case of six isodoublets with
$Y=\frac{1}{2}$, and set $\lambda_\phi=\lambda_H$ at $M_Z$~\footnote{This 
choice is motivated by the MSSM with its $({\tilde t}_L,
{\tilde b}_L), {\tilde t}_R$ and ${\tilde b}_R$. We have checked that
similar conclusions hold if we use just three `coloured' isodoublets.
The main features of Figs.~\ref{fig1} and \ref{fig2} are somewhat
accentuated, whereas Fig.~\ref{fig3} is essentially unchanged.}. We can
see
from Fig.~\ref{fig1} that, as one increases the coupling between $\phi$
and $H$, and hence the contribution $M_0$ to the scalar masses from the
coupling with the Higgs field, the critical scale $\mu_c$ at which the
Higgs coupling changes sign increases from its starting value of (just
below)  $10^6$ GeV, which is attained for small $M_0$. However, one can
see quite dramatically in Fig.~\ref{fig2} what happens when we try to
increase the value of $M_0$ so that this instability of the effective
potential does not occur below the Planck scale. The fine-tuning required
is so acute that, as one increases $M_0$ from 70.9 GeV to 71.0 GeV, the
behaviour of $\lambda_H(\mu)$ as a function of $\mu$ switches from one
that becomes negative at $\mu \, \sim \, 5 \times 10^7$ GeV to one that
reverses its fall just below that scale, and becomes so large at the
Planck scale that perturbative calculations become
unreliable.

The absence of new bosonic particles at LEP and the Tevatron indicates
that $M \gappeq 100$~GeV. On the other hand, it is clear that the value
of $M_0$ required to stabilize the effective potential must increase as
$M$ increases. However, $M$ must be less than $10^6$~GeV, since this is
the scale at which the instability appears in the Standard Model, and
$M_0$ is limited by the electroweak scale.
Since one would expect $M_0$ to be determined by the weak-interaction scale, 
 it is apparent from 
Fig.~\ref{fig1} that 
the scale $M$ at which new physics is introduced could not be much larger 
than $10^4$~GeV, and the fine-tuning visible in Fig.~\ref{fig2} is
exacerbated for larger $M$. For this reason, we would argue that $M
\lappeq 10^5$~GeV.

\section{Fine-Tuning}

Supersymmetry is the only known theory that provides
the requisite fine-tuning
in a natural way. In a supersymmetric theory, the quartic Higgs coupling
is forced
by supersymmetry to take the value
\beq \lambda_H \ = \ \frac{3}{4}(g^2+g^{\prime \, 2}) \label{susy1} \eeq 
at the tree level,
and its running is determined by the $\beta$ functions of the 
$SU(2)$ and $U(1)$ gauge couplings.

We can attempt to mimic
such a theory by starting from the Standard Model 
with a Higgs mass of 115 GeV, and using the R.G.E. of the Standard
Model to run the effective potential up to
1.4 TeV, which is the scale where
the Standard Model  $\lambda_H$ has the value specified by 
the relation (\ref{susy1}). It is at this point that we introduce six new
isodoublets of scalar particles that mimic the third generation
of (left- and right-handed) squarks. In order for the quartic Higgs
coupling to run as the required combination of gauge couplings, we need to
impose the following relation on $M_0$~\footnote{This awkward expression
arises because, in a genuine
supersymmetric model, the left- and right-handed squarks couple to the
Higgs field
in different ways, as reflected in the correct matter parts of the
$\beta$ functions for the gauge couplings. In our mimicry, we have
assumed for simplicity that all the scalar particles couple in the same
way.}:
\beq
M_0^4 \ = \ \frac{v^4}{8}\left\{ \left( g_t^2-\frac{(g^2+g^{\prime \, 2})}{8}
 \right)^2 \, + \,  
\left( \frac{135 \, g^4 +231 \, g^{\prime \, 4} - 18 g^2    
g^{\prime \, 2}}{576} \right) \right\}.
\label{M0formula}
\eeq
The important feature of this relation is the fact that, when substituted
into (\ref{dbetlam}), the terms proportional to the top-quark Yukawa
coupling
cancel those of (\ref{bethsm}). In order for this relation to be
maintained as the couplings run, we need to impose the condition
\beq
\lambda_\phi \ = \ \frac{g_t^2}{12} \, - \, \frac{4}{9} g_3^2,
\label{relation}
\eeq
where $g_3$ is the QCD coupling,
and we have dropped some terms proportional to powers of $g$ and
$g^\prime$, which have a negligible effect.

In contrast, in a supersymmetric model, the couplings are guaranteed to
run in the
required fixed ratio, because of contributions to the $\beta$ functions
from loops involving Higgsino and gaugino particles. However, in our
mimicry,
such extra fermions have not been added. The running
of the quartic Higgs coupling in a supersymmetric model receives a
contribution
\beq
\frac{1}{16 \pi^2} \left\{ - \, 3 \left( 5 g^4 \, + \,
 2 g^2 g^{\prime \, 2} +g^{\prime \, 4} \right) \, 
+ \, 2 \lambda_H \left( 3 g^2 + g^{\prime 2} \right) \right\}
\label{contribution}
\eeq
from these fermionic superpartners, whereas the 
supersymmetric one-loop $\beta$ function
for the top-quark Yukawa coupling~\cite{MV} has a contribution
\beq
\frac{3}{16\pi^2} g_t^3
\label{higgsinobit}
\eeq
from a loop involving a Higgsino and
\beq
\frac{8}{48\pi^2} g_t g_3^2
\label{gauginobit}
\eeq
from a loop involving a gluino.
Moreover, the $\beta$ functions for the gauge couplings~\cite{2loopjones}
also receive contributions from these fermionic superpartners.

\begin{figure}
\begin{center}
\leavevmode
\hbox{\epsfxsize=12 cm
\epsfbox{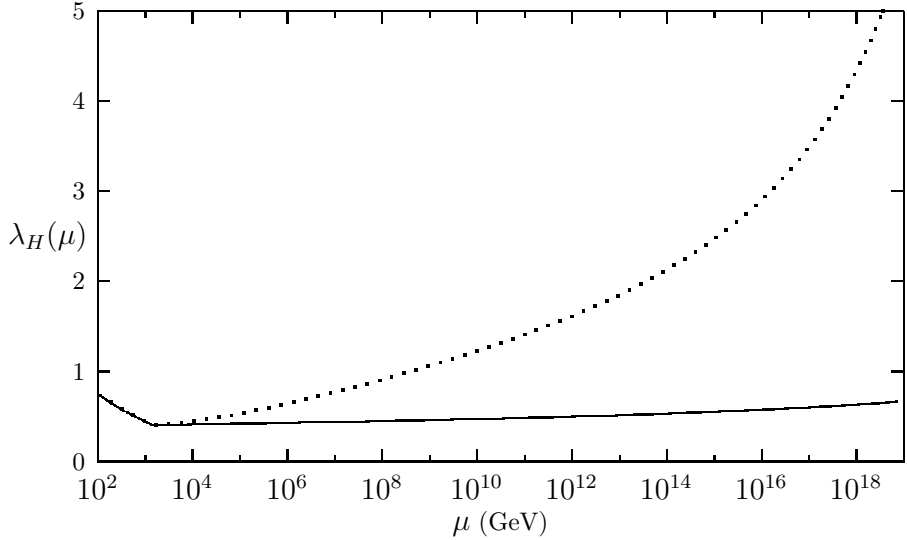}}
\vspace{1cm}
\caption{\it An example of the role played by fermionic superpartners
in the running of $\lambda_H$. The solid line corresponds to a genuine
supersymetric model, whereas the dotted line gives the running of the
quartic
Higgs coupling when the contributions from fermionic Higgsino 
and gaugino superpartners have been removed.}
\label{fig3}
\end{center}
\end{figure}

The essential role played by these fermionic superpartners can be seen
in Fig.~\ref{fig3}, in which we compare the running of $\lambda_H$
in the case of a genuine supersymmetric model with that in our mimic
model, that contains
the required number of extra scalar multiplets with couplings
tuned at the erstwhile supersymetric threshold to coincide with the
relations forced
by supersymmetry. In the genuine case, there is a very small rise in the
quartic coupling above the supersymmetric threshold, due to the fact
that
the $\beta$ functions for $g$ and $g^\prime$ have positive values.
On the other hand, we can see quite clearly that, when we remove
the contributions from the fermionic superpartners, we are again
faced
with the problem of a quartic coupling that rises too fast. 

We note that most of this effect is due to diagrams involving an internal
Higgsino, whereas those involving gauginos have only a small effect on the
running of $\lambda_H$.

\section{Electroweak Radiative Corrections}

A further constraint on the parameters involving additional scalar multiplets
arises from consideration of the precision weak parameters
\cite{PT} $S$ and $T$, that
characterize `oblique' weak corrections. 

We recall that scalar isospin multiplets with exact mass degeneracy, {\it
do not} make any contribution to these parameters. 
However, the $SU(2)$-breaking mass term $\propto (\Delta M^2)$ discussed
above {\it does}
introduce contributions to $S$ and $T$, which are given by
\beq \Delta S \ = \ =\frac{n  \, Y}{\pi} \sum_{i=-I}^I 
 i \ln (1+\eta i), \label{deltaS1}  \eeq

\bea \Delta T & = & \frac{n \, (\Delta M^2)}{8\pi M_W^2 \sin^2\theta_W} 
 \sum_{i=-I}^I \left( I(I+1)-i(i-1)\right) 
 \Bigg\{ \eta \ln(1+i \, \eta)+1+\eta(i-1/2)  \nonumber \\ & - & 
  \frac{\left(1+\eta(i-1)\right)^2}{\eta} 
\ln\left( \frac{1+\eta \, i}{1+\eta (i-1)}\right)
 \Bigg\}, \label{deltaT1} \eea
where $\eta=(\Delta M^2)/M^2$ is the fractional mass-splitting between
members of the multiplet.
These expressions simplify to
\beq \Delta S \ = \ \frac{n  \, Y}{9 \pi} \frac{(\Delta M^2)}{M^2} 
 I(I+1)(2I+1), \label{deltaS2}  \eeq
\beq \Delta T \ = \  \frac{n  \, Y}{6 \pi \sin^2 \theta_W }
 \frac{(\Delta M^2)}{M_W^2} 
 I(I+1)(2I+1), \label{deltaT2}  \eeq
in the limit $\eta \ll 1$.

The most stringent limit from precision electroweak data is that on the
quantity $T$. From the experimental error on this quantity
 we conclude that
any new physics can contribute at most $\pm 0.14$~\footnote{Note that,
since $(\Delta M^2)$ can take either sign, 
the contribution to $T$ can also take either sign. Thus, one could 
arrange a conspiracy
to produce cancellations with further new 
physics at some much higher scale, but here we neglect such a
possibility.}.
This means, for example, that if one were to add six 
isodoublets, the quantity
$(\Delta M^2)$ could be no greater than $400 \  {\mathrm GeV}^2$.
This implies, in terms of quartic
couplings, that the ratio of the $SU(2)-$breaking quartic coupling
to the Higgs quartic coupling $\lambda_H$ could be no more than
$0.04$~\footnote{Alternatively and equivalently, if $M_0 \sim m_t \sim
200$~GeV, one must enforce $(\Delta M^2) / M_0^2 \lappeq 0.01$.}.
Such an unnatural fine-tuning would not normally survive the higher-order
corrections due to $\lambda_H$. 

However, supersymmetry provides a natural suppression of this effect.
Provided that the soft supersymmetry-breaking mass is large compared with
the
contribution to the squark masses due to the Yukawa couplings, the squark
mass eigenstates have $SU(2)$ mass splittings which although considerably
larger than the above-mentioned limit, are almost equal and opposite, so
that the contribution to $T$ is comfortably small.

\section{Conclusions}

We have learnt in this paper that, if $m_H = 115$~GeV and we
want the effective Higgs potential to remain stable all the way up to the
Planck scale of $10^{19}$~GeV, the Standard Model must be supplemented by
a remarkably supersymmetric-seeming set of new physics. The top quark
must be accompanied by one or more scalar multiplets, whose couplings must
be tuned very finely if the effective potential is to steer a course
between the Scylla of a negative scalar coupling and the Charybdis of
non-perturbative couplings. This fine-tuning is provided in a
supersymmetric model by the stop and Higgsino supersymmetric partners
of the top quark and the Higgs boson, and the characteristic relations 
between couplings that they preserve. thus, a light Higgs could be
construed
as {\it prima facie} evidence for the supersymmetrization of the top quark
and the Higgs boson.

In order to extend the sell-by scale of the Standard Model far beyond
$10^6$~GeV, the parameters of the new physics must be quite finely tuned. 
Supersymmetry is one example of new physics that fulfils all the
requirements: any alternative should quack in a similar way. 

\noindent
{\bf Acknowledgement}\\
{~}\\
We thank Ben Allanach for helpful discussions.

\end{document}